\documentclass{iau} 
\usepackage{graphicx}

\title[Solar forcing of hydrological phenomena] 
{Solar activity forcing of \\ terrestrial hydrological phenomena}

\author[Pablo J.D. Mauas, Andrea P. Buccino \& Eduardo Flamenco]   
{Pablo J.D. Mauas$^1$, Andrea P. Buccino$^1$, \and Eduardo Flamenco$^2$}

\affiliation{$^1$Instituto de Astronom\'\i a y F\'\i sica del Espacio,
\\ Universidad de Buenos Aires, CONICET\\C.C. 67 Suc. 28 - 1428
\\Buenos Aires, Argentina 
\\ email: {\tt pablo@iafe.uba.ar, abuccino@iafe.uba.ar}
\\[\affilskip]
$^2$Instituto Nacional de Tecnolog\'\i a Agropecuaria, \\
Rivadavia 1439, 1033, \\
Buenos Aires, Argentina}

\pubyear{2017}
\volume{328}  
\jname{Living around Active Stars}
\editors{D. Nandi, A. Valio \& P. Petit, eds.}

\begin{document}

\maketitle

\begin{abstract}

Recently, the study of the influence of solar activity on the
Earth's climate received strong attention, mainly due to the
possibility, proposed by several authors, that global warming is not
anthropogenic, but is due  to an increase in solar
activity. Although this possibility has been ruled out, there are strong 
evidences that solar variability has an influence on Earth's climate, in 
regional scales.

Here we review some of these evidences, focusing in a particular
aspect of climate: atmospheric moisture and related quantities like
precipitation. In particular, we studied the influence of  
activity on South American precipitations during centuries. First, we
analyzed the stream flow of 
the Paran\'a and other rivers of the region, and found a very strong
correlation with Sunspot Number in decadal time scales. We found a
similar correlation between Sunspot Number and tree-ring
chronologies, which allows us to extend our study to
cover the last two centuries.

\end{abstract}


\section{Introduction}

\begin{figure}[t]
\begin{center}
 \includegraphics[width=3.8in]{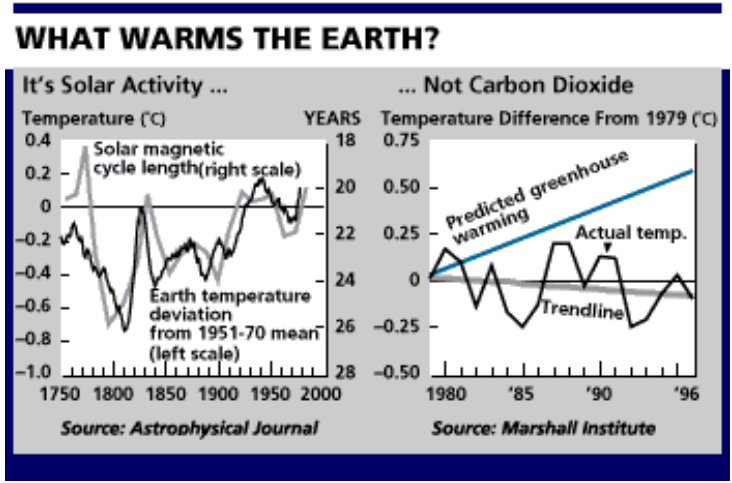} 
 \caption{Article on the Wall Street Journal (12/4/1997)}
   \label{spoken}
\end{center}
\end{figure}

In the last decades, several authors proposed that global warming is
not anthropogenic, but is due instead to an increase in solar
activity, a proposition which resulted in a strong interest to study 
the influence of solar activity on the Earth's climate. This
discussion was, of course, of great political interest,
and had a strong repercussion in the media. For example, on December
4, 1997, on the Wall Street Journal appeared an article on the subject
entitled ``Science Has Spoken: Global Warming Is a Myth'' (see
Fig. \ref{spoken}). This article, together with a copy of a
scientific-looking paper (of which there are three 
versions, e.g. Soon et al. 1999), was
massively sent to North American scientists, accompanied by a petition to be
presented to the Congress of the Unites States opposing the
ratification of the Kyoto protocol.  

This article was based on the results obtained by \cite{FCL91} and
\cite{LFC95}, who found a similarity between the length of the solar
cycle (LCS), smoothed with a 1-2-2-2-1 filter, and the 11-yr running mean of
the Northern Hemisphere temperature anomalies. However, this studies
were seriously objected by \cite{LG00} and \cite {Laut03}. In
particular, these results were obtained using the actual,
non-smoothed, LSCs for the last 4 cycles. Using the right values,
already available 10 years later, it can be seen that the solar cycles
had approximately the
same length than in the 1970s, while the temperature continued to
increase (e.g. see \cite{DP05}).

Several years later, \cite{FCS97} and \cite{sven98} found that total
cloud cover changed in phase with the flux of galactic cosmic
rays (GCR), which are modulated by the interplanetary magnetic field
associated with the solar wind and, therefore, with solar activity.
They proposed a mechanism for the influence of solar activity on
climate, in which GCR would affect cloud formation on Earth, through
ionization of the atmosphere. Therefore, during periods of higher solar
activity, when the interplanetary magnetic field is larger, and
therefore less GCR hit the Earth, the cloud cover would be smaller.

Later on the observed 
agreement was lost, although \cite{MS00} proposed that it was still
visible with low clouds. This theory was criticized by different
reasons (e.g. Laut 2003), in particular because GCR should affect more
strongly high clouds than low ones. Furthermore, \cite{UC01} studied
observations from the ground obtained at 90 meteorological stations
in the US during more than 90 years, and found the opposite
correlation. At present, the correlation found by Svensmark and
collaborators cannot be seen in the data. In fact, \cite{LF07} found
that all possible solar forcings of climate had trends opposite to
those needed to account for the rise in temperatures measured in the
last century.

Moreover, the idea that the Sun has played a significant role in
modern climate warming was mainly based on a general consensus 
that solar activity has been increasing during the last 300 years, 
after the Maunder Minimum, with a maximum in the late 20th
century, which some researchers called the Modern Grand
Maximum. However, this increase in solar activity has been identified
as an error in the calibration of the Group Sunspot Number. When this
error is corrected, solar activity appears to have been relatively
stable since the end of the Maunder Minimum (see e.g. \cite{sval12}, and
the official IAU release
\footnote{https://www.iau.org/news/pressreleases/detail/iau1508/}). 

However, even if global warming cannot be attributed to an increase in
solar activity, there is strong evidence that activity can influence
terrestrial climate, in local scales. In what follows we will review
some of that evidence, in particular the one referred to hydrological
phenomena, and review our recent work on the subject.

\section{Solar activity and hydrological phenomena}

Usually, studies focusing on the influence of solar activity on
climate have concentrated on Northern Hemisphere temperature or sea
surface temperature. However, climate is a very
complex system, involving many other important
variables. Recently, several studies have focused in a different
aspect of climate: atmospheric moisture and related quantities like,
for example, precipitation.

Perhaps the most studied case is the Asian monsoon, where
correlations between precipitations and  solar activity have been found
in several time scales. For example, \cite{2001Natur.411..290N}
studied the monsoon in Oman between 9 and 6 kyr ago, and found
strong coherence with solar variability. \cite{2002E&PSL.198..521A}
found that the
monsoon intensity in India followed the variations of the solar irradiance  on 
centennial time scales during the last
millennium. \cite{2003Sci...300.1737F} studied  the
Indian monsoon during the Holocene, and found that intervals of weak  solar
activity correlate with periods of low monsoon precipitation, and viceversa.
On shorter time scales, \cite{1997GeoRL..24..159M}, found that, at
multi\-de\-ca\-dal time scales, when solar irradiance is
above normal there is a stronger correlation between the El Ni\~no 3
index and the monsoon rainfall, and
viceversa. \cite{2005GeoRL..3205813B} and \cite{2004GeoRL..3124209K},
among others, also found correlations between solar activity and
Indian monsoon in decadal time scales.

The monsoon in southern China over the past 9000 years was studied by
\cite{2005Sci...308..854W} who found that higher solar
irradiance corresponds to stronger monsoon. They proposed that the monsoon
responds almost immediately to the solar forcing by rapid atmospheric
responses to solar changes.

\cite{TiRa14} studied groundwater recharge rates in the Chinese region of
Mongolia. Groundwater recharge is the hydrologic process where water
moves downward from surface water to an aquifer. They found a strong
stationary power at 200-220 years, significant at more than 95\%
confidence level, with wet periods coincident with strong solar activity
periods.

All these studies found a positive correlation, with periods of
higher solar activity corresponding to periods of larger
precipitation. In contrast, \cite{2001E&PSL.185..111H} studied a
6000-year record of precipitation and drought in northeastern China,
and found that most of the dry periods agree  with stronger
solar activity and viceversa. In the American continent, droughts in
the Yucatan Peninsula have been associated with periods of strong
solar activity and have even been proposed to cause the decline of
the Mayan civilization (\cite{2001E&PSL.192..109H}).

\begin{figure}[b!]
\begin{center}
 \includegraphics[width=4.in]{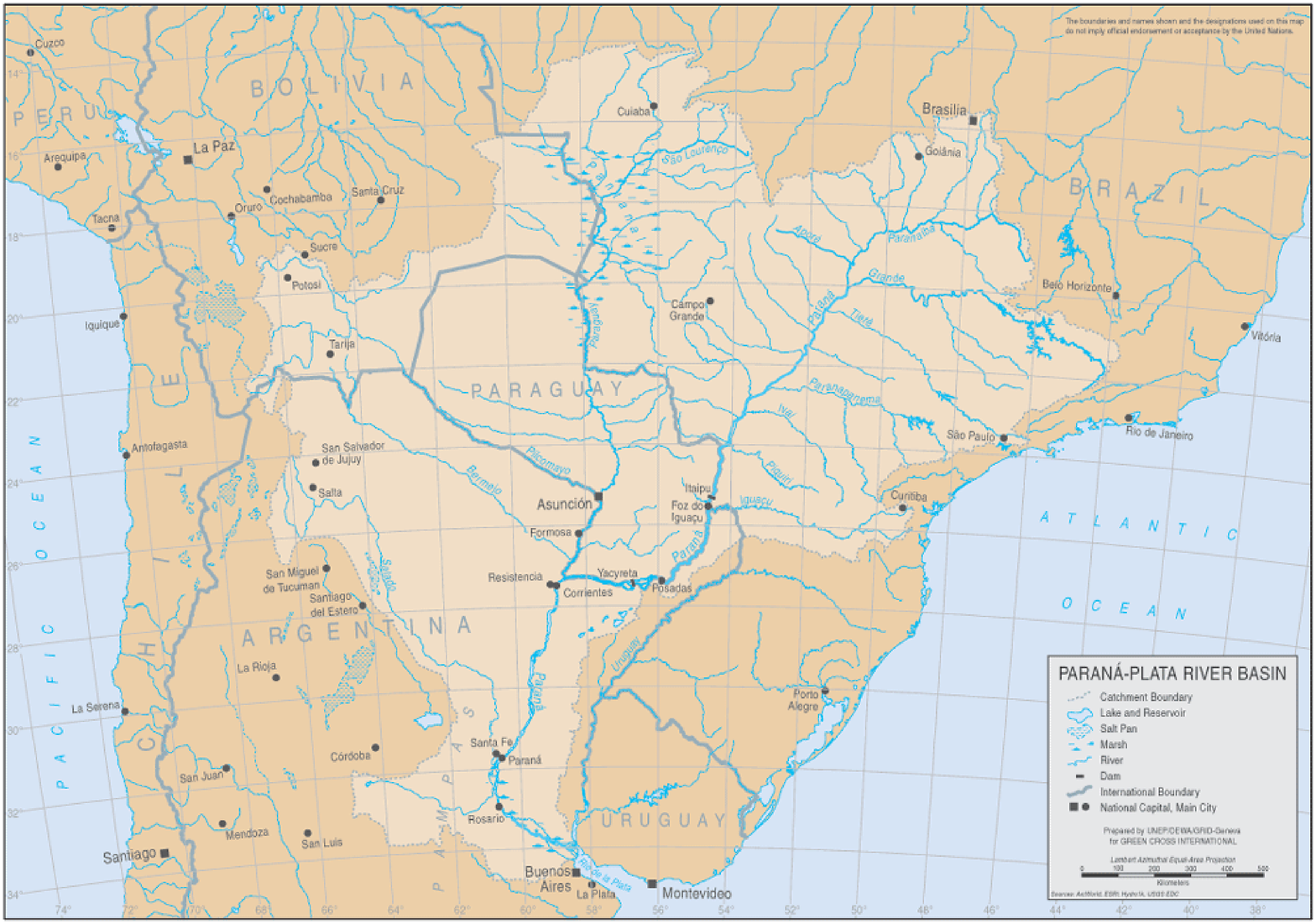} 
 \caption{The Paran\'a basin}
   \label{parana-mapa}
\end{center}
\end{figure}

In the same sense, studies of the water level of the East African
Lakes Naivasha (\cite{2000Natur.403..410V}) and Victoria
(\cite{stager05}), found that severe droughts were coincident with
phases of high solar activity and that rains increased during periods
of low solar irradiation. To explain these differences it has been
proposed that in equatorial regions enhanced solar irradiation causes
more evaporation increasing the net transport of moisture flux to the
Indian region via monsoon winds (Agnihotri \etal\ 2002).

However, these relationships seem to have changed sign around 200
years ago, when strong droughts took place over much of tropical Africa
during the Dalton  minimum, around 1800-1820 (\cite{stager05}).
Furthermore, recent water levels in Lake Victoria were studied by
\cite{2007JGRD..11215106S}, who found that during the 20th century,
maxima of the $\sim$11-year sunspot cycle were coincident with water
level peaks caused by 
positive rainfall anomalies $\sim$1 years before solar maxima. These
same patterns were also observed in at least five other East African lakes,
hinting that these relationships between sunspot number and rainfall were 
regional in scale.

In \cite{2005MmSAI..76.1002M} we took a different approach, and we
proposed to use the stream flow of a large river, the Paran\'a in
southern South America, to study precipitations in a large area (see
below). In this direction, \cite{Ruzmaikin2006} found signals of solar
activity in the river Nile using spectral analysis techniques. They
reported an 88-year variation present both in solar 
variability and in the Nile data. \cite{2008JGRD..11312102Z} studied
the stream flow of the Po river, and found a correlation with
variations in solar activity, on decadal time scales.


\section{Stream flow of the Paran\'a River}

River stream flows are
excellent climatic indicators, and those with continental scale basins
smooth out local variations, and can be particularly useful to study
global forcing mechanisms. In particular, the Paran\'a River
originates in the southernmost part of the Amazon forest, and it flows
south collecting water from the countries of Brazil, Paraguay,
Bolivia, Uruguay, and Argentina (see Fig. \ref{parana-mapa}).
It has a basin area of over 3.100.000
km$^2$ and a mean stream flow of 20.600 m$^3$/s, which makes it the
fifth river of the world according to drainage area and the fourth
according to stream flow. 

\begin{figure}[t]
\begin{center}
 \includegraphics[width=4.4in]{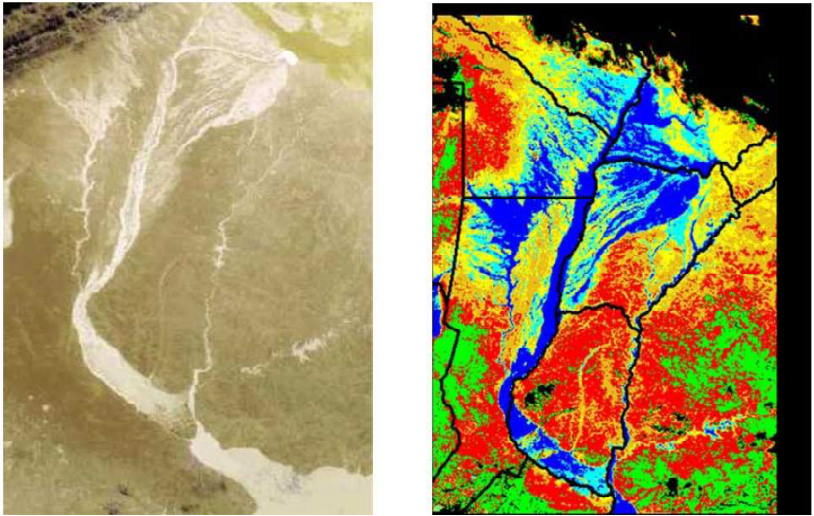} 
\vspace*{-2.0 cm}
 \caption{Left: Image taken with the AVHRR instrument on board a NOAA satellite,
showing an area of 1200 km x 500 km, during the flood of the Paran\'a of 1997-1998.
Right: An image in false colors, with political divisions over-imposed. The main
area is Argentina. To the East, Uruguay and Brazil. To the North, Paraguay.}
   \label{parana-flood}
\end{center}
\end{figure}

\begin{figure}[t!]
\begin{center}
 \includegraphics[width=3.in]{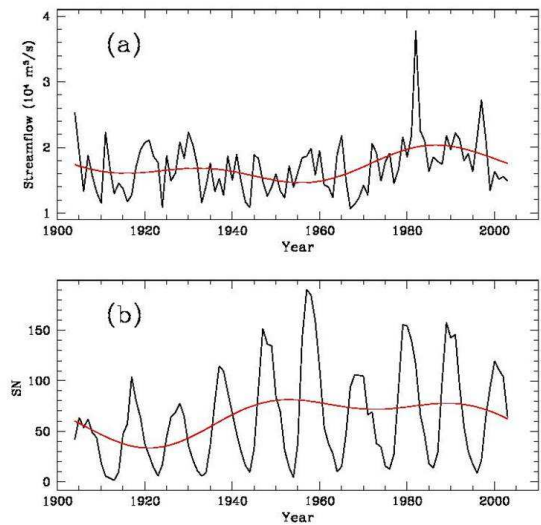} 
 \caption{a) Paran\'a's annual stream flow at the Corrientes
gauging station. (b) Yearly international sunspot number (SN).
The secular trends, obtained with a low-pass Fourier filter with a
50-yr cutoff, are shown as thick lines.  }
   \label{parana-datos}
\end{center}
\end{figure}

\begin{figure}[t]
\begin{center}
 \includegraphics[width=3.in]{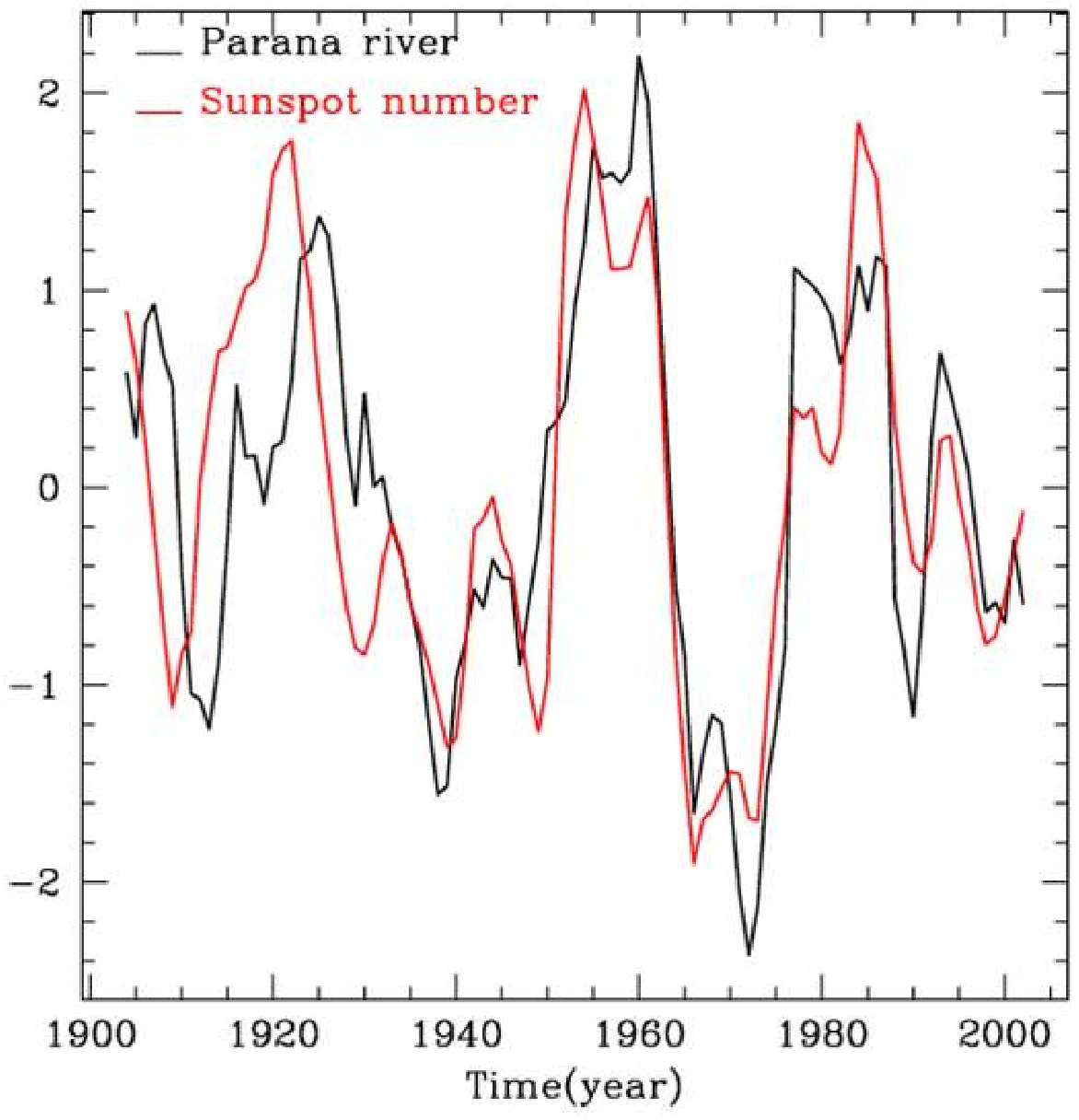} 
 \caption{The detrended time series for the Paran\'a's stream flow  and the
Sunspot Number. The detrended series were obtained by subtracting
from each data series the corresponding secular trend and were smoothed by an
11-yr-running mean to eliminate the solar cycle. Both series were standardized by
subtracting the mean and dividing by the standard deviation, to avoid introducing
arbitrary free parameters. The Pearson’s correlation coefficient is R=0.78.}
\label{parana-correl}
\end{center}
\end{figure}

Understanding the different factors that have an impact on the flow of
these rivers it is fundamental for different social
and economic reasons, from planning of agricultural or hydroenergetic
conditions to the prediction of floods and
droughts. In particular, floods of the Paran\'a can occupy very large
areas, as can be seen in Fig. \ref{parana-flood}. During  the  last
flood,  in  1997, 180 000 km$^2$ of land were covered with water, 125 000
people had to be evacuated, and 25 people died. Together, 
the three largest floods of the Paran\'a during the 20th
century caused economic losses of five billion dollars.

In \cite{MFB08} we studied the stream flow data measured at a
gauging station located in the city of Corrientes, 900 km
north of the outlet of the Paran\'a. It is measured continuously from
1904, on a daily basis. The yearly data are shown in Fig. \ref{parana-datos}
together with the yearly sunspot number (SN), which we use as a
solar-activity indicator. Also shown in the figure are the trends,
obtained with a low-pass Fourier filter with a 50 years cutoff. 

In Fig. \ref{parana-correl} we show the stream flow and the SN
together. In both cases we have subtracted the secular trend shown in
Fig. \ref{parana-datos} from the annual data, and we have performed an
11 yr running-mean to smooth out the solar cycle. 
We have also normalized both quantities by subtracting the mean and
dividing by the standard deviation of each series. These lasts steps
have been done to avoid introducing two free parameters, the relative
scales and the offset between both quantities. 

It can be seen that there is a remarkable visual agreement between the
Paran\'a's stream flow and the sunspot number. In fact, the Pearson's
correlation coefficient is $r=0.78$, with a significance level,
obtained through a t-student test, higher than 99.99\%. It can also be
noted that in this area wetter conditions  coincide with periods of
higher solar activity. 

A few years later, in \cite{MFB11} we found that the correlation still held
when more years of data were added. In particular, between
1995 and 2003 the Paran\'a's stream flow and the mean Sunspot Number
have both decreased by similar proportions. This is of particular
interest, since Solar Cycle 23 was the weakest since the 1970s: SN
for the years 2008 (2.9) and 2009 (3.1) have been the lowest since
1913, and the beginning of Solar Cycle 24 was delayed
by a minimum with the largest number of spotless days since the
1910s. At the same time, the mean
levels of the Paran\'a discharge were also the lowest since the 1970s
(see Fig. \ref{parana-correl}).


\section{Other South-American Rivers}

\begin{figure}[t]
\begin{center}
 \includegraphics[width=3.in]{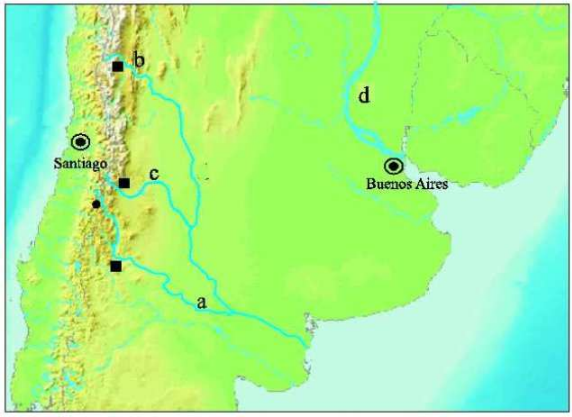} 
 \caption{Colorado hydrologic system. The rivers we studied are 
a. Colorado, b. San Juan, c. Atuel and d. the lower part of the Paran\'a river. The
stream flow (squares) and snow (dot) measuring stations are also indicated.}
\label{otros-rios-mapa}
\end{center}
\end{figure}

In \cite{MFB11} we followed up on the study of the influence of solar
activity on the flow of South American rivers. In that paper we
studied the stream flow of the Colorado river, and two of its tributaries,
the San Juan and the Atuel rivers. We also analyzed snow
levels, measured near the sources of the Colorado (see
Fig. \ref{otros-rios-mapa}).

The Colorado river marks the north boundary of the Argentine Patagonia,
separating it from the Pampas, to the northeast, and the Andean region of
Cuyo, to the Northwest. Its origin is on the eastern slopes of the
Andes Mountains, from where it flows southeast until it discharges in
the Atlantic Ocean. 
The Atuel, which originates in the glacial Atuel Lake, at 3250 m above
sea level in the Andes range, and the 500 km long San Juan river, join the
Colorado downstream of it’s gauging station. Therefore, the data given by the
three series are not directly related. 

\begin{figure}[b]
\begin{center}
 \includegraphics[width=5.5in]{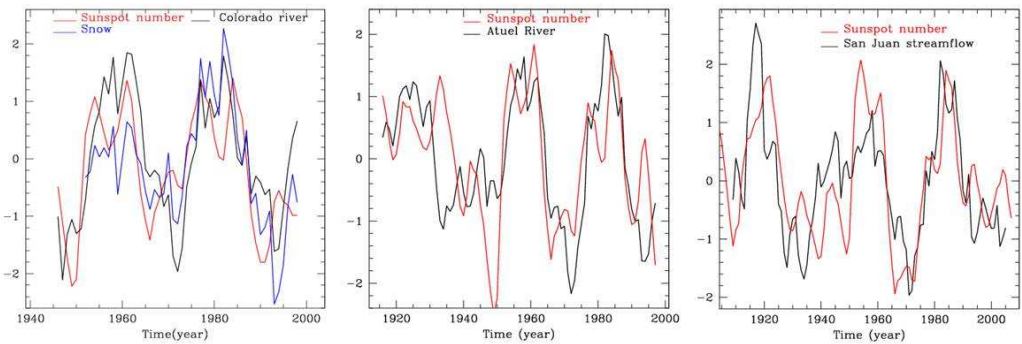} 
 \caption{The detrended and renormalized stream flows compared with Sunspot Num\-ber. In
   the left panel the snow level is also shown.} 
   \label{otros-rios-corr}
\end{center}
\end{figure}

Unlike the Paran\'a, whose stream flow is directly related to precipitation,
the regime of all these rivers is dominated by snow melting, and their stream
flows reflect precipitation accumulated during the winter, and melted during
spring and summer. To directly study the snow precipitation, we
complete our data with measurements of the height of snow accumulated
in the Andes at 2250 m above Sea level, close to the origin of the
Colorado (see Fig. \ref{otros-rios-mapa}), which were measured in situ
at the end of the winter since 1952. In fact, the correlation between
the stream flow of the Colorado and the snow height is very good, with
a correlation coefficient $r=0.87$, significant to a 99\% level.

In Fig. \ref{otros-rios-corr} we compare the multidecadal component of
the stream flows with the corresponding series for the sunspot
number. In all cases 
we proceed as with the data in Fig. \ref{parana-datos}:
we smoothed out the solar cycle with an 11-year running mean, we
detrended the series by subtracting the long term component, and
we standardized the data by subtracting the mean and dividing by the
standard deviation. In the panel corresponding to the Colorado, we
also include the snow height.

It can be seen that in all cases the agreement is remarkable. The correlation
coefficients are 0.59, 0.47, 0.67 and 0.69 for the Colorado, the snow level,
the San Juan and the Atuel, respectively, all significant to the
96-97\% level. Although all these rivers have maximum stream flow
during Summer, there is a 
big difference, however, between the regimes of the Paran\'a and the remaining
rivers: for these ones, the important factor is the intensity of the precipitation
occurring as snow during the winter months, from June to August. For
the Paran\'a, what is  
most important is the level of the precipitation during the summer
months. It should also be noticed that, here again, stronger
activity coincides with larger precipitation.


\section{Tree rings}

Tree rings are the most numerous and widely
distributed high-resolution climate archi\-ves in South America. During
the last decades, variations of temperature, stream flow, rainfall and
snow were reconstructed using tree-ring chronologies from subtropical
and temperate forests, which are based on ring width, density and
stable isotopes (see \cite{Boninsegna2009210} and references therein).  

\begin{figure}[b]
\begin{center}
 \includegraphics[width=3.4in]{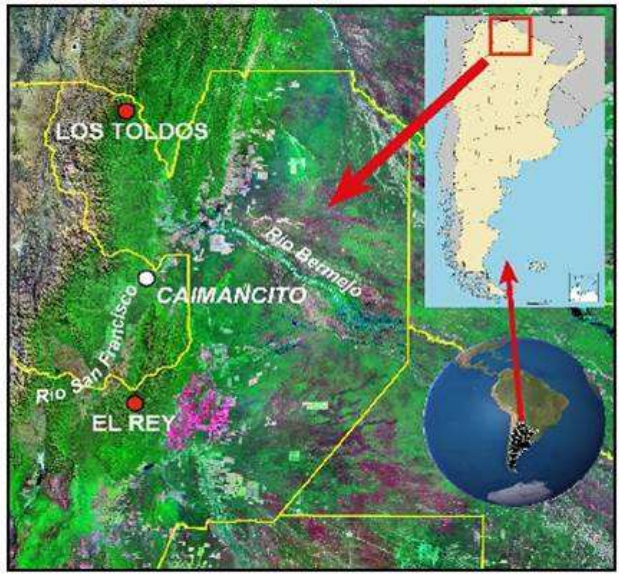} 
\caption{NorthWestern Argentina. Marked in red are the
locations from where we obtained the tree-ring chronologies. Image
acquired by the LANDSAT satellite.}
   \label{tr-mapa}
\end{center}
\end{figure}

\cite{villalba92} studied the spatial patterns of climate and
 tree-growth anomalies in the forests of northwestern
 Argentina. The tree-ring data set consisted of seven chronologies developed
 from Juglans and Cedrela (see Fig. \ref{tr-mapa}). They show that
 tree-ring widths in subtropical Argentina are affected by weather
 conditions from late winter to early summer. Tree-ring patterns
 mainly reflect the direct effects of the principal types of 
rainfall-patterns observed. One of these patterns is related to
precipitation anomalies concentrated in the northeastern part of the region.

\begin{figure}[t]
\begin{center}
 \includegraphics[width=3.6in]{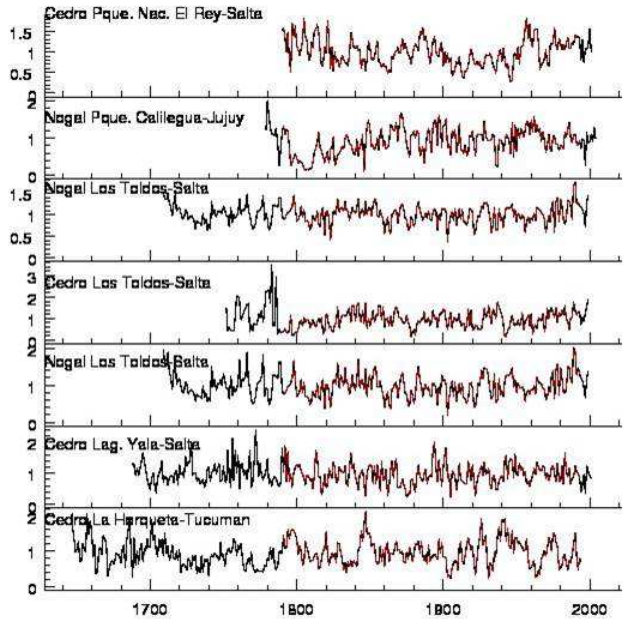} 
 \caption{Tree-rings chronologies from Cedrela and Junglans from
  NorthWestern Argentina. Marked in red are the data used in this 
  work.} 
  \label{tr-data}
\end{center}
\end{figure}

To extend back in time and to a larger geographical area the results
obtained previously, we study the relation between the Sunspot Number
and the tree-ring chronologies studied by \cite{villalba92}. These
data-sets are shown in Fig. \ref{tr-data}. It can be seen that the
shortest series starts in 1797, while the longest one goes back to the
XVI century. Here  we study only the data from 1797, where all the
series overlap.

\begin{figure}[t]
\begin{center}
 \includegraphics[width=3.4in]{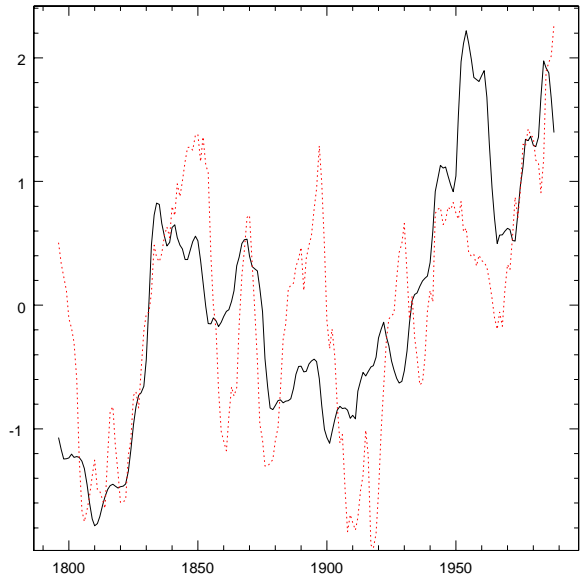} 
 \caption{Composite of the tree-rings chronologies (dashed-line) and Sunspot
   Number (\hbox{full}-line) smoothed by an 11-yr running mean to smooth out the
   solar cycle. Both series were normalized by subtracting the mean
   and dividing by the standard deviation. The correlation coefficient
   is r=0.69.}
   \label{tr-correl}
\end{center}
\end{figure}

The individual sets respond to local conditions, in the particular
location of the studied tree. To obtain an indication of global
conditions in the region, we built an index in the following
way. First, we shifted in time each tree-ring series to obtain the best
correlation with the Paran\'a's stream flow. In particular, in 1982 
and 1997 there were two very large annual discharges of the Paran\'a 
that are associated with two exceptional El Ni\~no events (see
Fig. \ref{parana-datos}). These two events, although weaker, can be
seen in the individual tree-ring series, with a small delay, different
in each case. We therefore built a composite series as the average
between each individual chronology, shifted to match the Paran\'a's
discharge. Finally, we took the 11-years running-mean, and normalized
the series as in the previous cases.

The resulting index is shown in Fig. \ref{tr-correl}, together with 
the Sunspot Number. It can be seen that also here the agreement is
quite good. The Pearson's correlation coefficient between both series
is R=0.69.

\section{Discussion}

Although the theory that Global Warming is caused by an increase in
Solar activity has been dismissed, particularly because activity and
temperature do not have similar trends anymore, it gave a strong
impulse to the studies on the relation between climate and
activity. In particular, there is strong evidence that the Sun could
have an influence in different climatic variables, in different
regions of the globe, and not always in the same sense. In particular,
we reviewed different studies which concentrate on different aspects
of atmospheric moisture, which in some regions reported positive
correlations, with stronger activity related to stronger
precipitations, and in others the opposite correlation, with strong droughts
coincident with solar activity maxima. There are also regions of the
world where this relation changed signs over time.

In particular, we studied different climatic indicators in southern
South America. First, we concentrated in the stream flow of one of the
largest rivers of the world, the Paran\'a. We found a strong
correlation on decadal time scales between the river's discharge and
Sunspot Number. We later found that this correlation was still present
during the large solar minimum between Cycles 23 and 24, which
corresponded to a period of very low flows in the Paran\'a. 

We can also find in historical records this coincidence between periods
with smaller solar activity and low Paran\'a's discharge. In
particular, during the period known as the Little Ice Age there are
different reports pointing out to low discharges. For example, a
traveler of that period mentions in his diary that in 1752 the level
of the river was so small that the small ships of that time could not
navigate it. At present, the river can be navigated as far north as
Asunci\'on in Paraguay  by ships 4 times larger (\cite{Iri99}). There
other climatic records which point out to reduced precipitations in this
region during the  Little Ice Age (see \cite{piovano09} and references
therein). It is well known that the  Little Ice Age was coincident
with the Maunder Minimum, and was  perhaps 
caused by low solar activity (e.g. \cite{1976Sci...192.1189E}).

To check if the solar influence is also present in other areas
of  South America, we studied the flow of three other rivers of the
region, and the snow level from a mountain-high station of the same
area. Also in this cases we found  a strong correlation between 
the Sunspot Number and the stream flows, after removing the
secular trends and smoothing out the solar cycle.

Finally, to extended both the area coverage and the temporal
baseline, we studied a composite of seven tree-ring chronologies
affected by precipitations, starting at the end of the XVIII century.
Also in this case we found the same correlation with Sunspot Number.

We point out that, in all cases, we found a correlation in the
intermediate time scale. We removed the secular trends when present
(e.g, for the Paran\'a and the Sunspot Number), which are not
correlated. We also smoothed out the solar cycle, since on the yearly
timescale, the dominant factor influencing precipitations is El
Ni\~no. The results we found show that {\it 
decades} of larger precipitations correspond to  {\it decades} of higher
activity, with these variations overimposed on the corresponding
secular trends.

In all cases, the correlation we found is positive, i.e., higher
precipitations correspond to larger solar activity, in a very large
area. 

Since another mechanism that has been proposed to explain the
Sun-Earth connection involve the modulation of Galactic Cosmic Rays,
we also studied the correlation between the Paran\'a's discharge and
two other solar-activity indexes: the neutron count at Climax,
Colorado, available since 1953, and the aa index, which is an
indication of the disturbance level of the magnetic field of the Earth
based on magnetometer observations of two stations in  England and
Australia, which is available since 1868. Both indexes can be used to
test the GCR hypothesis.

We found that the Paran\'a's stream flow is correlated with both
neutron count and the aa index. This was expected, since all activity
indexes are correlated among them.  However, the correlation with
Sunspot Number was strongest, suggesting a direct link between solar
irradiance and precipitations. 

It has been shown that variations in solar insolation affect the
position of the Inter Tropical Convergence Zone (ITCZ)
(\cite{2004GeoRL..3112214P}, \cite{2001Sci...293.1304H}). 
\cite{2006GeoRL..3319710N} proposed that a displacement southwards of
the ITCZ would enhance precipitations in the tropical regions of
southern South America. We found that the increase in precipitations
are seen both in the Southern Hemisphere's summer when the
ITCZ is over the equator, close to where the Paran\'a has its origin,
and during winter, when the ITCZ moves north, and precipitations increase
further South.

\end{document}